\documentclass[apjl]{emulateapj}

\shorttitle{CRISP Spectropolarimetric Imaging of Penumbral Fine Structure}
\shortauthors{Scharmer et al.}

\begin{document}

\title{CRISP Spectropolarimetric Imaging of Penumbral Fine Structure}

\author{G.B. Scharmer\altaffilmark{1}, G. Narayan\altaffilmark{1,2},
  T. Hillberg\altaffilmark{1,2}, J. de la Cruz
  Rodriguez\altaffilmark{1}, M.G. L\"ofdahl\altaffilmark{1},
  D. Kiselman\altaffilmark{1}, \\P. S\"utterlin\altaffilmark{1}, 
  M. van Noort\altaffilmark{1}, A. Lagg\altaffilmark{3}}
\altaffiltext{1}{Institute for Solar Physics, Royal Swedish Academy of Sciences,
  AlbaNova University Center, SE-106\,91 Stockholm, Sweden}
\altaffiltext{2} {Stockholm Observatory, Stockholm University}
\altaffiltext{3}{Max Planck Institute for Solar System Research,
Max-Planck-Str. 2, DE-37191 Katlenburg-Lindau, Germany}

\begin{abstract}
  We discuss penumbral fine structure in a small part of a pore,
  observed with the CRISP imaging spectropolarimeter at the Swedish
  1-m Solar Telescope (SST), close to its diffraction limit of
  0\farcs16. Milne--Eddington inversions applied to these Stokes data
  reveal large variations of field strength and inclination angle over
  dark-cored penumbral intrusions and a dark-cored light bridge. The
  mid-outer part of this penumbra structure shows $\sim$0\farcs3 wide
  spines, separated by $\sim$1\farcs6 (1200~km) and associated with
  30\degr{} inclination variations.  Between these spines, there are
  no small-scale magnetic structures that easily can be be identified
  with individual flux tubes. A structure with nearly 10\degr{} more vertical
  and \emph{weaker} magnetic field is seen midways between two
  spines. This structure is co-spatial with the brightest penumbral
  filament, possibly indicating the location of a convective upflow
  from below.
\end{abstract}


\section{Introduction}

The discovery of dark cores in sunspot penumbral filaments (Scharmer
et al.\@ 2002) suggests that the basic elements of penumbral fine
structures are observable.
The nature of individual dark cores was investigated with
high-resolution multi-line spectra, but without polarization
information, by Bellot Rubio et al.\@ (2005). They concluded that the
cores are associated with 
weaker magnetic field strength, by 100--300~G\@.
Langhans et al.\@ (2007) measured the azimuthal variation of circular
polarization signal 
in the 6302~{\AA} \ion{Fe}{1} line 
of regular spots at different heliocentric distances. They found
that dark cores are associated with a \emph{strongly}
reduced field strength and a magnetic field that is more horizontal
than for their lateral brightenings by about 10\degr{}--15\degr{}.
Analysis of spectropolarimetric data from the Japanese satellite
Hinode also shows lower field strength in the dark cores, but only by
100--150~G, and small inclination changes of about 4\degr{}
(Bellot Rubio et al.\@ 2007).

We describe the first spectropolarimetric
observations with CRISP, an imaging spectropolarimeter built for the 
Swedish 1-m Solar Telescope (SST). The large, unobscured aperture of the 
SST corresponds to a diffraction-limited resolution that is twice as high 
as that of Hinode. 
We present CRISP observations of penumbral fine structure made at a spatial 
resolution close to the SST diffraction limit of 0\farcs16.
Using Milne--Eddington (ME) inversions applied to these data, we discuss
spatial variations of the magnetic field and line-of-sight (LOS) velocities
for penumbral structure seen over parts of a large pore.

\section{Observational data and processing}

CRISP is a spectropolarimeter,
based on a dual Fabry--P\'erot interferometer (FPI) system similar to
that described by Scharmer (2006). The spectral resolution is modest
(about 60~m{\AA} at 6302~{\AA}), to limit the number of wavelengths
needed without aliasing and to allow a large field-of-view (FOV) at
high image quality.
The present image scale is 0\farcs071/pixel. Further information about
CRISP will be given in a future publication. 

For polarimetric analysis, a polarizing beam splitter close to the
final focal plane is used with two 1k$\times$1k-pixel synchronized
CCD's and nematic liquid crystals (LC's). For image restoration, a
third CCD simultaneously records a wide-band image through the pre-filter of
the FPI system. 

\begin{figure*}
  \hfill
  \includegraphics[width=0.368\textwidth]{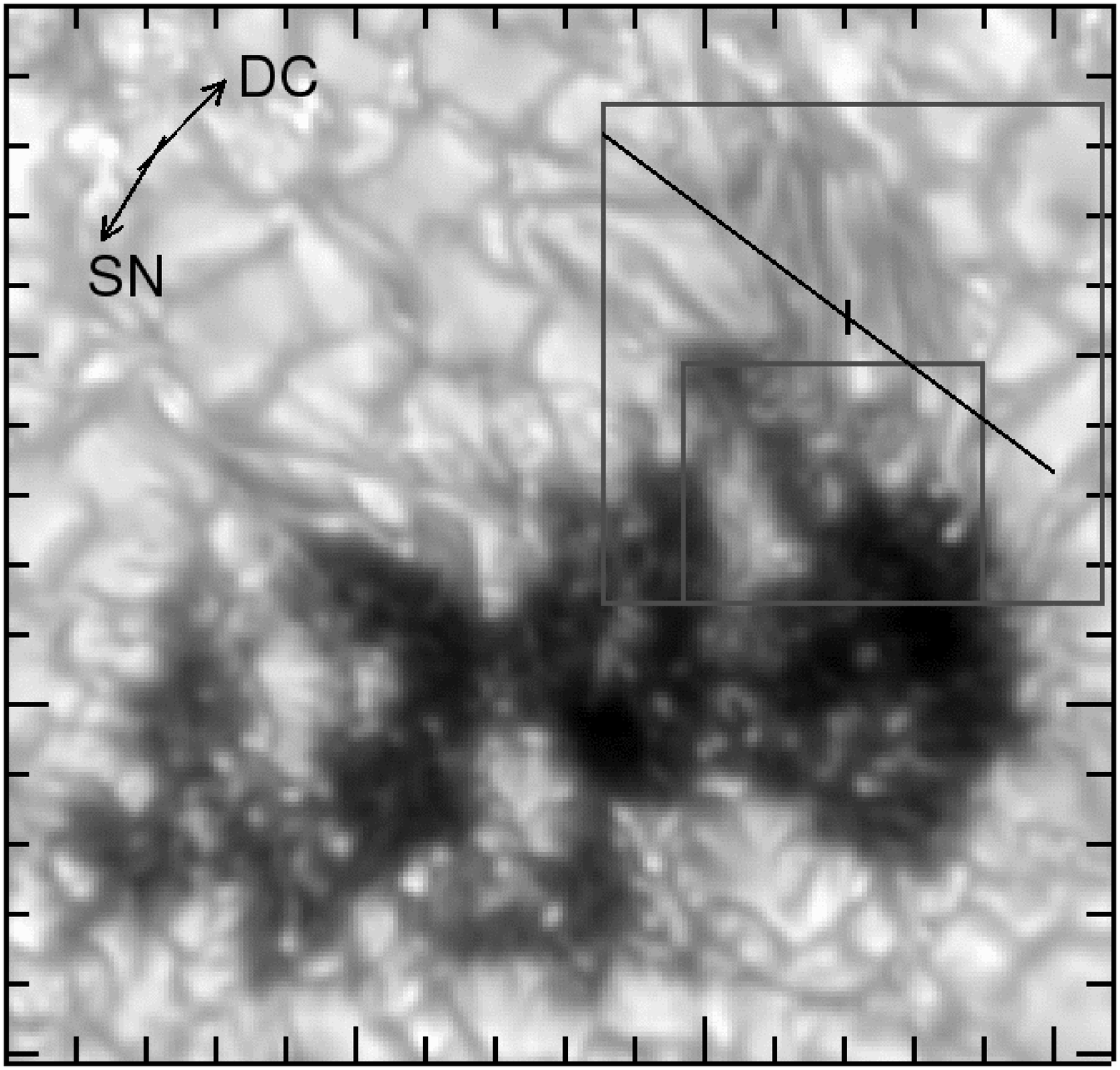}
  \hfill
  \includegraphics[bb=14 17 468 283,width=0.60\textwidth]{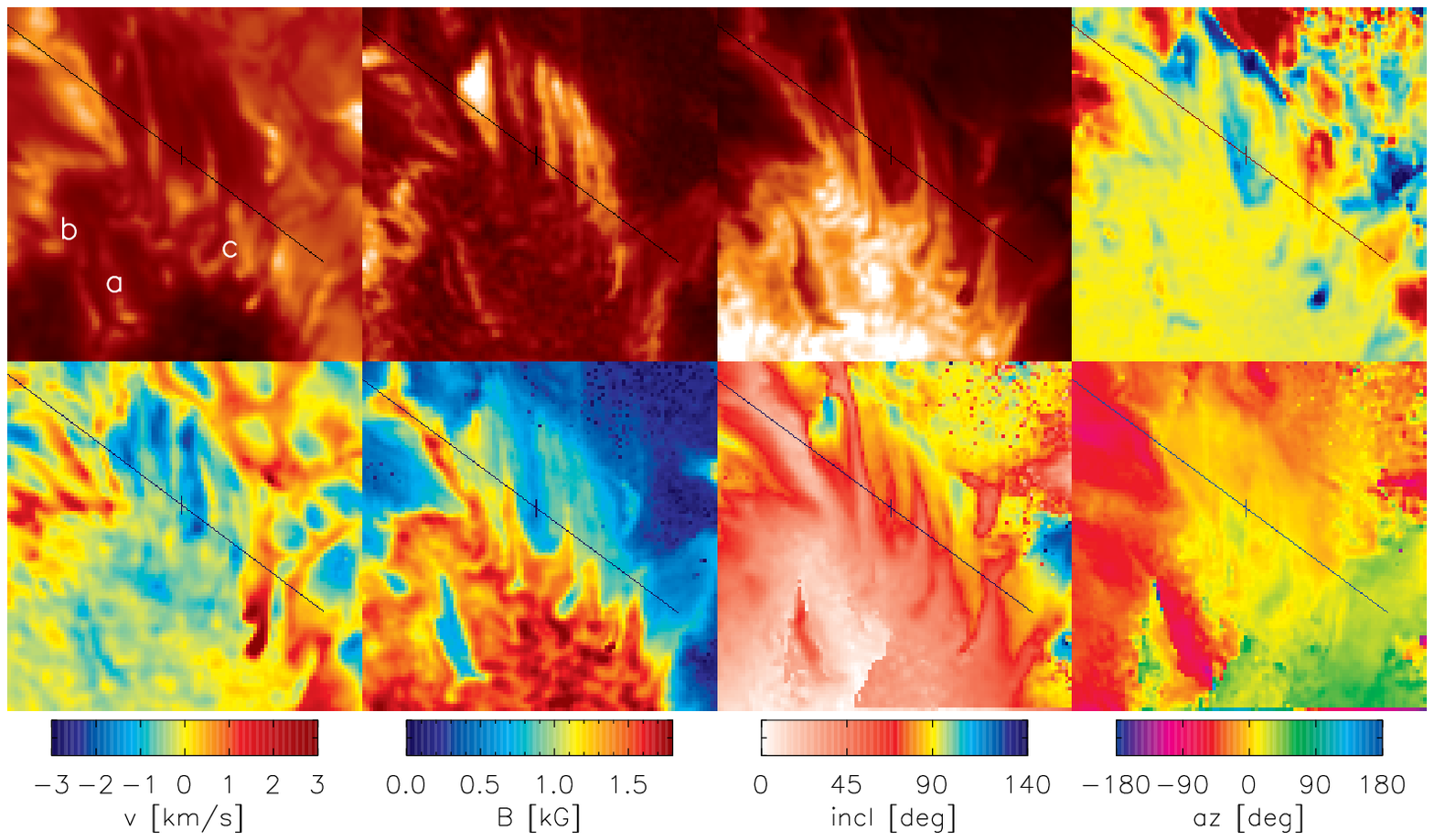}
  \hfill
  \caption{\textbf{Left:} The continuum image of one of the two pores
    observed with CRISP. The upper-right part of this pore and the
    associated light-bridge structure and penumbral filaments, indicated with the
    large box, are discussed in this Letter. The directions of solar disk center
    (DC) and solar north (SN) are indicated with arrows. Tick marks are
    separated by 1\arcsec. \textbf{Right:} The top row shows Stokes I at line
    center, the integrated linear and circular polarization and the Stokes V
    asymmetry (AA). The color coding of AA is such that yellow indicates
    small AA, green-blue negative AA (Stokes V is negative in the blue wing
    over most of the spot) and red positive AA.
    The bottom row shows the LOS velocity, field
    strength, inclination and azimuth angles derived from the inversions.
    The dark line indicates the cut along which plots are shown in Fig~\ref{gapflow}.
    The size of the boxes are $\sim 7.1\times 7.1$ arcsec.
    }
  \label{context}
  \label{quantities}
\end{figure*}

The data discussed in this Letter were recorded on 22 April 2008 at
approximately 11:58 UT. The target was two large pores of opposite
polarities (AR10992), located at approximately N18\,E4, corresponding
to a heliocentric distance of approximately 19\degr{}
($\mu=0.94$). We discuss data from a
small part of one of these pores, indicated by the larger of the two boxes
shown in Fig.\@~\ref{context}.

\begin{figure}
  \plotone{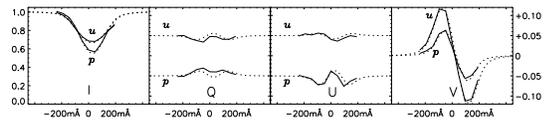}
  \caption{Two sets of observed (full) and fitted (dotted) Stokes
    profiles from the umbra (u) and penumbra (p). Stokes Q, U and V are
    normalized to the continuum intensity and plotted on the same scale,
    but with Q and U shifted vertically for clarity.}
  \label{profiles}
\end{figure}

The images recorded correspond to complete Stokes measurements at 11
line positions in steps of 48~m{\AA}, from $-240$ to
$+240$ m{\AA}, in the \ion{Fe}{1} line at 6302~{\AA}. In addition,
images were recorded at one continuum wavelength. The frame rate was
36~Hz, with an exposure time of 16~ms and a CCD readout time of
10~ms. For each wavelength and LC state, 14 images were
recorded per camera. Each sequence processed consists of about 630
images per CCD (1890 images in total) and was recorded during a total
time interval of 22~s. The images, covering a FOV of
$71\arcsec\times71\arcsec$ were divided into overlapping
$128\times128$-pixel subfields and all images from each subfield were
processed with MOMFBD image restoration techniques (van Noort et
al.\@ 2005). To further reduce noise, two consecutive 
sets of images were aligned and co-added. These images were
demodulated with respect to the polarimeter (pixel by pixel) and the
telescope polarization model developed by Selbing (2005) -- for
details, see also van Noort \& Rouppe van der Voort (2008). In
addition, the Stokes images were corrected for remaining I to Q, U and
V cross-talk with the aid of the Stokes images recorded in the
continuum.  The final Stokes images thus obtained were combined to
Stokes spectra. It was verified that nearly all Stokes V profiles are
normal with two lobes and small asymmetries, justifying the use of ME
inversion techniques with this data.
Stokes Q, U and V images were further inspected for evidence
of cross-talk from V to Q or U but several examples of strong V
features without co-spatial Q or U features seen in the 2D maps
suggest that such cross-talks must be small.
We finally estimated the noise level to be approximately
$1.0\cdot10^{-3}$ for Stokes V and $1.1\cdot10^{-3}$ for Stokes Q and
U.
For the analysis presented here, no attempt was made to compensate for
the telluric blend, instead the Stokes data obtained at
$+240$ m{\AA} were ignored. Also, any FOV variations of the
\emph{shape} of the FPI transmission profile from cavity errors in the
two etalons were ignored and the theoretical profile based on known
cavity separations and reflectivities was used. Examples of observed 
and the corresponding calculated Stokes profiles are shown in Fig.\@~\ref{profiles}.

The final Stokes images were processed with the ME inversion software
HELIX, developed by Lagg et al.\@ (2004). The following parameters
were determined by the code: LOS velocity, Doppler width, the gradient
of the source function and the field strength $B$, inclination and azimuth
angle of the magnetic field plus a parameter allowing adjustment of the continuum
level. Macroturbulence was set to zero. The ratio of line center 
to continuum opacity, $\eta_0$, was set to 16, the damping parameter to 1 
and the filling factor $f$ to 1. Because of the assumed unit filling factor, 
the derived magnetic field parameters represent locally averaged 
values, also along the LOS. The filling factor was fixed at unity for 
comparison with the inversions of Bellot Rubio et al. (2007), discussed in 
Sect. 3.2. Tests with free $f$ showed a few localized areas with 
strongly reduced $f$ and strongly increased $B$ but the average field 
strength, $B f$, was consistently close to what was obtained with $f=1$. 
Allowing $\eta_0$ to vary gave large changes in the source function gradient 
but had small effect on the magnetic field. These tests and tests with other
inversion codes convinced us that the inversion results are robust. 

Finally, we used the flat-field
images, recorded at steps of 24~m{\AA} and based on images accumulated
during 40~min, to produce a 6302 \ion{Fe}{1} line etalon cavity map,
corresponding to the Doppler shift bias introduced by the cavity
errors. This bias was subtracted from the Doppler shifts derived by
the inversion software. After this correction, the average Doppler
shifts of the two pores differed by about $60$~m\,s$^{-1}$. The average 
of the two pore Doppler shifts were used as reference for all velocities
measured. We used this calibration to calculate the
average convective blue shift of the part of the surrounding quiet sun
that showed no magnetic field significantly above the noise level. The
convective blueshift thus obtained was $210$~m\,s$^{-1}$. This is
close to the value obtained for this line from convection simulations
at a heliocentric distance of 19\degr, $-236$~m\,s$^{-1}$ by de la 
Cruz Rodriguez (in prep.), but differing by $130$~m\,s$^{-1}$ from the 
value obtained by Dom{\'{\i}}nguez Cerde{\~n}a et al. (2006).

\section{Results}

Figure~\ref{quantities} (right panel) shows input data and results of
the inversions. The total linear and circular polarizations are
calculated as $\int {(Q^2+U^2)^{1/2}} d\lambda/I_c $ and
$\int {| V |}d\lambda/I_c $, where $I_c$ is the continuum intensity,
and the Stokes V area asymmetry as $\int {V}d\lambda/\int {| V |}d\lambda$.

\begin{figure}
  \plotone{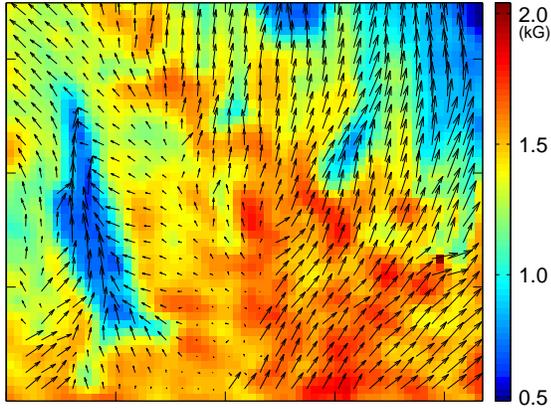}
  \caption{The horizontal magnetic field component
    within the small box  shown in Fig.\@~\ref{context}, overlaid on the field
    strength map. Large variations in the azimuthal angle are associated
    with the light bridge and the most prominent dark core associated with
    penumbral filaments. Tickmarks are separated by 1\arcsec.}
  \label{field}
\end{figure}

\subsection{Dark-cored structures}

A short irregular light bridge, labeled `a', and four dark-cored 
intrusions are seen in the pore. The dark cores
are barely visible in the continuum Stokes I image but clearly visible
in line center Stokes I, the integrated circular
polarization and the field strength maps. The light bridge dark core
`a' and the two strongest dark-cored intrusions, labeled `b' and `c', show
clear variations in the inclination and azimuth angle. The light
bridge dark core `a' and one of the other dark cores split up in Y-shapes
at their innermost parts, similar to what was first reported by
Scharmer et al.\@ (2002) and also seen in umbral dot simulations by
Sch\"ussler \& V\"ogler (2006). Figure~\ref{field} shows the horizontal magnetic 
field component for every second pixel within the small box in Fig.~\ref{quantities}, 
after resolving the 180\degr{} ambiguity and transforming the magnetic 
field to the solar local reference frame. 
The light-bridge dark core `a' and the strongest dark-cored
intrusion `c' are associated with a magnetic field that is more
horizontal, weaker and aligned with the main axis of the dark
cores. The minimum field strength in these dark cores is approximately
600--800~G, as compared to the average field strength in the
pore of 1600~G and a peak field strength of 2100~G. Crossing the
dark core of the light bridge (`a'), the inclination angle varies by
over 50\degr{} and by approximately 15\degr{} when crossing the dark core of
`c'. However, for neither of these dark cores is the magnetic field
close to horizontal. For dark cores `a', `b' and `c', the maximum
inclinations are 70\degr{}, 60\degr{}, and 55\degr, respectively. Figure~\ref{field}
shows large variations, by nearly 90\degr, in the magnetic field
azimuth angle across the light bridge dark core and 45\degr{}
variations across the dark core of `c'. The magnetic
field for both these dark cores is aligned with the main axis of the
cores.  The Doppler map shows blueshifts of 0.6~km\,s$^{-1}$ in the
light bridge structure, and about 0.3~km\,s$^{-1}$ along the core. 

\begin{figure}
  \includegraphics[bb=9 6 470 228, clip, width=\linewidth] {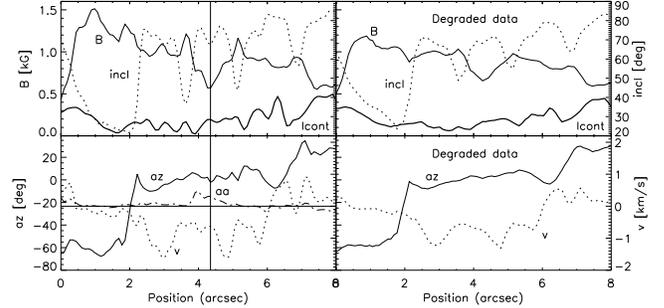}
  \caption{The inclination angle, azimuth
    angle, LOS velocity, field strength, continuum intensity and Stokes V
    area asymmetry (AA) along
    the line shown in Fig.\@~\ref{context}, with the mark midways
    corresponding to the vertical line shown in the present Figure.
    AA uses the same scaling as the LOS velocity in the
    RHS of the figure.
    \textbf{Left:} Inversions based on SST data. \textbf{Right:} Inversions
    based on SST
    data degraded to a resolution corresponding to a 50-cm telescope with
    34\% linear obscuration.}
  \label{gapflow}
\end{figure}

\subsection{Penumbra spine structures}

The penumbral filamentary structures seen further away from the center of the pore
show only a single thin dark core.  Other filamentary structures are
similar to those of the outer parts of mature
penumbrae. Figure~\ref{gapflow} shows plots of the inclination angle,
azimuth angle, LOS velocity, field strength, continuum intensity 
and Stokes V area asymmetry (AA, labeled `aa' in the plot) along the thin 
black line in Fig.\@~\ref{quantities}.
As can be seen, AA is small except close to one of the spines. The 
most prominent variations are
those of the inclination, varying by up to 40\degr{} over distances of
only 0\farcs25--0\farcs3. In contrast, the azimuth angle varies by
only $\pm7\degr$. The plot and Fig.\@~\ref{quantities} also show
spines (dark structures in the inclination map), first identified by
Lites et al.\@ (1993), associated with small, on the order of 150~G,
enhancements of the field strength in the inner and mid parts of the
spines. Along the spines, the field strength and inclination variations
appear as radial extensions of the stronger and more vertical field
of the umbral part of the pore. The FWHM of the spines, as measured
from the inclination map, is in the range 0\farcs25--0\farcs35.
In the outer parts of this penumbra, the field strength of
the spines is very close to that of their surroundings.  This field
strength variation with radial distance is
qualitatively similar to that reported by Westendorp Plaza et al.\@
(2001), derived from ASP data obtained at much lower spatial
resolution. The LOS velocity shows large variations.  Where the
magnetic field is more horizontal, the LOS blueshift is around
1.2--1.7~km\,s$^{-1}$ and where it is more vertical, it is
approximately 0.3--0.4~km\,s$^{-1}$. Figs.\@~\ref{quantities} and
\ref{gapflow} show a weak spine-like structure, indicated with a short
dark vertical line in Fig.\@~\ref{quantities}, in the inclination map.
Here the magnetic field is more vertical than in the immediate surroundings,
but only by less than 10\degr, and the field strength reaches a local
\emph{minimum} in contrast to what is the case for the two spines 
surrounding this structure. This feature can be seen clearly as a 
`gap' in the integrated linear polarization map shown in Fig.~\ref{quantities} and it 
shows up in all ME inversions made, including those with free filling 
factor. Thus we see spine-like
structures associated with a more vertical field of two types:
Those that are associated with stronger field in the mid-inner
penumbra and those that are associated with weaker field throughout
the penumbra. Similar conclusions were first drawn by Westendorp Plaza
et al.\@ (2001), c.f. their Fig.\@~11. However, the present data show
no obvious evidence of field lines spreading in the azimuthal 
direction away from the spines, as found for filaments close to the 
symmetry axis on the limb side by Borrero et al.\@ (2008). 
Along the structure with minimum field strength, the LOS blueshift is 
strongly reduced to only about $0.3$~km\,s$^{-1}$. This region of minimum 
field strength corresponds to the brightest penumbral filament seen in the 
continuum image.

Another notable feature is the variation of the inclination along the edge
outlining the outer penumbral boundary.  Along this edge the
inversions return inclinations close to 90\degr, with some local inclinations 
as large as 100\degr, suggesting field lines that return down into the
photosphere at shallow angle with respect to the surface. Redshifts in 
the range 0.6--1.0~km\,s$^{-1}$ are seen here. Several examples of
abnormal (single- or multilobed) Stokes V profiles are found within this region.
A localized, very strong redshift, most likely to be a downflow, can be seen 
associated with a narrow filamentary intrusion in the lower-right part of
Fig.\@~\ref{quantities}. At the location of the peak redshift, the
Stokes profiles are sufficiently shifted in wavelength that the inversions
cannot be trusted. Here, Stokes V profiles are strongly asymmmetric or 
abnormal with one or three lobes. It is likely that the LOS velocity 
exceeds 5~km\,s$^{-1}$ in this region. Similar strong downflows at the edge of 
umbrae were recently reported by Shimizu et al. (2008).

The spatial variations in the inferred properties of the magnetic field are 
in qualitative agreement with findings from inversions of Hinode Stokes 
data (Bellot Rubio et al. 2007), but at the high spatial resolution of the 
present data, some of these variations are found to be significantly larger. 
We have estimated the effect of higher spatial resolution by degrading the 
resolution of the present SST data to that of a 50~cm telescope with a linear 
central obscuration of 34\%. We binned the data over 2$\times$2 pixels 
to a pixel size of 0\farcs14. This process gives an impression of the 
improvement gained by the larger aperture of the SST over a 50-cm telescope 
such as Hinode/SOT. Inversions based on this degraded data are shown in the 
right panel of Fig.~\ref{gapflow}. This demonstrates a reduction of the 
inclination variations across the spines of about 50\%. 
Small-scale variations in field strength and LOS velocity are reduced by
more than 50\% but large-scale variations remain intact, suggesting (as
confirmed by other tests discussed above) that the inversions are robust.
The minimum field strength for the light bridge is increased only marginally,
but for dark cores `b' and `c', the increase is from 720 to 980~G and 
from 780 to 1040~G respectively. The maximum inclination angles are decreased
from 70\degr{} to 50\degr{} (`a'), from 60\degr{} to 48\degr{} (`b') and 
from 55\degr{} to 50\degr{} (`c'). 
This strongly suggests that the inferred stronger variations in magnetic field
properties across spines, a dark-cored light bridge and dark-cored
filaments, compared to what is obtained from Hinode data (Bellot Rubio et
al. 2007), are to a large extent due to the higher spatial resolution of the 
present data.

\section{Discussion}

We have presented ME inversions based on data obtained with the CRISP
imaging spectropolarimeter, used with the 1-m SST. The spatial
resolution of this Stokes data represents a break-through in
ground-based spectropolarimetry and a major improvement also as
compared to recent Hinode data. The large variations in field strength and 
inclination angle inferred from the present data can to a large extent can 
be explained with the high spatial resolution of the SST/CRISP data.

The penumbra observed is partial, covering only a small part of the
pore observed. We have analyzed dark-cored filamentary 
structures intruding into this pore and a lightbridge-like 
structure, detached from the surrounding photosphere.
The three dark-cored structures analyzed are associated with strongly
reduced field strength (around 50\% relative to their surroundings),
and a significantly more horizontal magnetic field (by 15\degr--50\degr) 
than outside the dark cores. The variations in field strength and
inclination across the dark-cored filament are consistent with analysis of
earlier SST magnetogram (Stokes V) data (Langhans et al. 2007). Even with the
high spatial resolution of that and the present SST data, the magnetic field is
found to be far from horizontal above the penumbral dark cores.

The inclination map shows a pronounced `spine' structure (Lites et al.\@ 1993)
with a magnetic field that is locally more vertical by
$\sim 30\degr$ and that, except in the outer penumbra, is locally stronger
by about 150~G\@. Within these spines, the LOS velocities are strongly reduced.
The spines seen in Fig.\@~\ref{context} are separated by about 1200~km.
Such widely separated spines were evident also in earlier SST magnetogram
data, discussed by Scharmer et al. (2007). We do not find any evidence for
horizontal flux tubes with diameters in the range 100-250~km modeled in
numerous papers (e.g., Borrero et al. 2007; Tritschler et al. 2007; Ruiz Cobo
\& Bellot Rubio 2008), in the outer penumbra. Midways between two of the spines,
a faint spine-like structure is seen in the inclination map. The locally weaker
and more \emph{vertical} magnetic field of this structure is in contradiction with
an interpretation in terms of a horizontal flux tube. The region with weaker field strength
coincides with the brightest penumbral filament seen in the continuum. A possible
interpretation is that the faint spine-like structure is related to a weak convective 
upflow, making the magnetic field overlying
that upflow locally more vertical. This interpretation must be regarded as
speculative in view of the small LOS velocities measured. Future CRISP
observations of larger sunspots are likely to clarify this and also to provide
more critical constraints on models.

\acknowledgements 
The Swedish 1-m Solar Telescope is operated on the
island of La Palma by the Institute for Solar Physics of the Royal
Swedish Academy of Sciences in the Spanish Observatorio del Roque de
los Muchachos of the Instituto de Astrof{\'\i}sica de Canarias. CRISP was
funded by the Marianne and Marcus Wallenberg Foundation. JdlCR was supported 
by a Marie Curie Early Stage Research Training Fellowship of the European 
Community's Sixth Framework Programme under contract number 
MEST-CT-2005-020395: The USO-SP International Graduate School for Solar Physics.

{\it Facilities:} \facility{SST (CRISP)}

\end{document}